\documentclass[aps,pra,twocolumn,floatfix]{revtex4-1}

\pdfoutput=1

\usepackage[utf8]{inputenc}
\usepackage{amssymb,amsmath}
\usepackage{bm}
\usepackage{graphicx}
\usepackage[usenames,dvipsnames]{xcolor}
\usepackage[colorlinks,bookmarks=false,citecolor=blue,linkcolor=cyan,urlcolor=blue]{hyperref}

\newcommand{\vac}{|\text{vac}\rangle}
\renewcommand{\ss}{\text{ss}}
\newcommand{\Z}{\ensuremath{\mathbb{Z}}}

\begin{document}

\title{Quantum noise in a transversely pumped cavity Bose--Hubbard model}
\author{D\'avid Nagy}
\affiliation{Institute for Solid State Physics and Optics, Wigner Research Centre for Physics, Hungarian Academy of Sciences, H-1525 Budapest P.O. Box 49, Hungary}
\author{G\'abor K\'onya}
\affiliation{Institute for Solid State Physics and Optics, Wigner Research Centre for Physics, Hungarian Academy of Sciences, H-1525 Budapest P.O. Box 49, Hungary}
\author{Peter Domokos}
\affiliation{Institute for Solid State Physics and Optics, Wigner Research Centre for Physics, Hungarian Academy of Sciences, H-1525 Budapest P.O. Box 49, Hungary}
\author{Gergely Szirmai}
\affiliation{Institute for Solid State Physics and Optics, Wigner Research Centre for Physics, Hungarian Academy of Sciences, H-1525 Budapest P.O. Box 49, Hungary}

\begin{abstract}
We investigate the quantum measurement noise effects on the dynamics of an atomic Bose lattice gas inside an optical resonator. We describe the dynamics by means of a hybrid model consisting of a Bose--Hubbard Hamiltonian for the atoms and a Heisenberg--Langevin equation for the lossy cavity field mode.  We assume that the atoms are prepared initially in the ground state of the lattice Hamiltonian and then start to interact with the cavity mode. 
We show that the cavity field fluctuations originating from the dissipative outcoupling of photons from the resonator lead to vastly different effects in the different possible ground state phases, i.e., the superfluid, the supersolid, the Mott- and the charge-density-wave phases. In the former two phases with the presence of a superfluid wavefunction, the quantum measurement noise appears as a driving term leading to depletion of the ground state. The time scale for the system to leave the ground state is presented in a simple analytical form. For the latter two incompressible phases, the quantum noise results in the fluctuation of the chemical potential. We derive an analytical expression for the corresponding broadening of the quasiparticle resonances.

\end{abstract}

\maketitle

\section{Introduction}
\label{sec:intro}

Cavity Quantum Electrodynamics (CQED) is devoted to studying the
interaction of the electromagnetic field with atoms under the best
possible control of circumstances. The radiation field is tailored
by resonators in order to select spatially and spectrally one or a few
relevant field modes. The atomic positions are controlled as well;
either by well-defined trajectories across the resonator, or by 
trapping the atoms in well-defined positions \cite{Nussmann2005Submicron,khudaverdyan2008controlled}. The control over the components
allows for accessing the coherent quantum dynamics of the coupled
atom-field system. The first milestone has been the demonstration of
the vacuum Rabi splitting \cite{thompson1992observation}, which is the benchmark of strong
coupling between the induced electric dipole of the atom and the
cavity mode. The possibility of observing coherent processes, such as the
Rabi oscillation, is limited by the spontaneous photon scattering into
modes other than the cavity mode. The larger the electric dipole
coupling strength $g$ with respect to the atomic spontaneous emission
rate $\gamma$ and the cavity mode linewidth $\kappa$, the shorter is the dynamical timescale needed to resolve the Jaynes--Cummings spectrum. Various nonlinear quantum effects, e.g., the photon blockade
\cite{Hamsen2017TwoPhoton}, or the two-photon gateway \cite{Kubanek2008TwoPhoton} have been observed in the strong coupling regime of cavity QED.  

With the use of ultracold atoms in CQED experiments, the magnitude of the collective
coupling of an atomic ensemble to the cavity mode can be significantly
enhanced to higher values than those characterising the loss rates. This allows for applying
large detuning between all laser excitations and the internal
atomic resonances, and thereby the atomic scattering loss processes
can be significantly suppressed. As a consequence,  the atom-cavity
dynamics can be controlled on much longer time scales. This opportunity opened up the way to a new regime of cavity QED experiments, where the
spatial motion of an atom cloud couples coherently to the dynamics
of the cavity field mode \cite{Ritsch2013Cold}.  The effective Hamiltonian
describing the system includes characteristic frequencies well below
the single-atom coupling strength $g$, e.g., the so-called recoil
frequency $\omega_R=\hbar{}k^2/(2m)$, where $k$ is the cavity mode wavenumber and $m$ is the atomic mass. These experiments revealed, for example, a
Dicke-type superradiant phase transition of an atomic superfluid in
the cavity
\cite{Baumann2010Dicke,Baumann2011Exploring,Klinder2015Dynamical}, and
demonstrated optomechanical strong coupling between vibration and
field intensity \cite{Brennecke2008Cavity}. The limitation of the coherent dynamics on the time scales longer than the inverse of the recoil frequency  originates from the spontaneous photon scattering again, however, (i) the spontaneous emission from atoms into free-space modes is strongly suppressed, (ii) the photon loss from the cavity mode into external modes is weakly coupled into the dynamics in a rather indirect way. 

A very new generation of CQED experiments
\cite{Landig2016Quantum,Klinder2015Observation} introduces a time scale which is
even longer than the inverse recoil frequency. Ultracold atoms trapped
in optical lattices sustained by the cavity can tunnel between
adjacent sites. The many-body quantum state of the atoms in the
lattice sites becomes relevant to the dynamics of the cavity field amplitude \cite{Mekhov2012Quantum,Elliott2015Multipartite,kozlowski2017quantum}. The tunneling
time depends on the depth of the trapping potential which is a novel
control parameter in the cavity QED system. The effective Hamiltonian corresponds to the family of bosonic Hubbard-type lattice models
extended to include the cavity field mode
\cite{FernandezVidal2010Quantum}. In these Hamiltonian systems exotic
new phases of lattice bosons appear due to the cavity mediated
global-range interactions
\cite{Li2013Latticesupersolid,Dogra2016Phase,Chen2016Quantum,Niederle2016Ultracold,Sundar2016Lattice,Panas2017Spectral,Flottat2017Phase,Liao2018Theoretical,blass2017quantum,Caballero2016Quantum,DallaTorre2016Dicke,Zheng2018Anomalous}. In fermionic lattices, cavity-induced topologically nontrivial \cite{Kollath2016Ultracold,Sheikhan2016Cavityinduced,Mivehvar2017Superradiant} states can be generated. All these research topics stimulate nowadays a significant theoretical and experimental interest in atomic lattice gases integrated in CQED systems.

Despite both the pronounced theoretical and experimental interest, very little effort has been made to study the limitations of a Hamiltonian approach. In a very recent paper Chiacchio and Nunnenkamp studied the time evolution of the density matrix with integrating the master equation  \cite{Chiacchio2018Tuning}. They found that in the bad cavity limit, which is close to the experimental situation of Ref. \cite{Landig2016Quantum}, the steady state is an infinite temperature state. In the experiments, however, they find a more or less coherent evolution during the time of the measurements. 

In this paper we investigate the time limitation on the coherence in cavity Bose--Hubbard models. We consider the cavity photon loss as the dominant dissipative process. Since the outcoupled photons can be directed to a photo-detector, the fluctuations associated with the loss process can be equally well conceived as the measurement-induced backaction on the quantum system. We derive simple analytical formulae in terms of experimentally measurable quantities to quantify the time limitation of the Hamiltonian description. In Section~\ref{sec:model}, we introduce one example of a cavity Bose--Hubbard model and derive the Heisenberg--Langevin equations that take into account the fluctuations of the cavity mode. In Sec.~\ref{sec:effmodel}, we derive an effective model for the lattice bosons by adiabatically eliminating the cavity field. In Sec.~\ref{sec:noiseeff}, we calculate and compare the effects of dissipation noise in the (i) superfluid-type and (ii) in the Mott-type phases. Finally, we summarise the results in Sec.~\ref{sec:sum}.

\section{Cavity Bose--Hubbard model}
\label{sec:model}

\begin{figure}[tb!]
\centering
\includegraphics[width=\columnwidth]{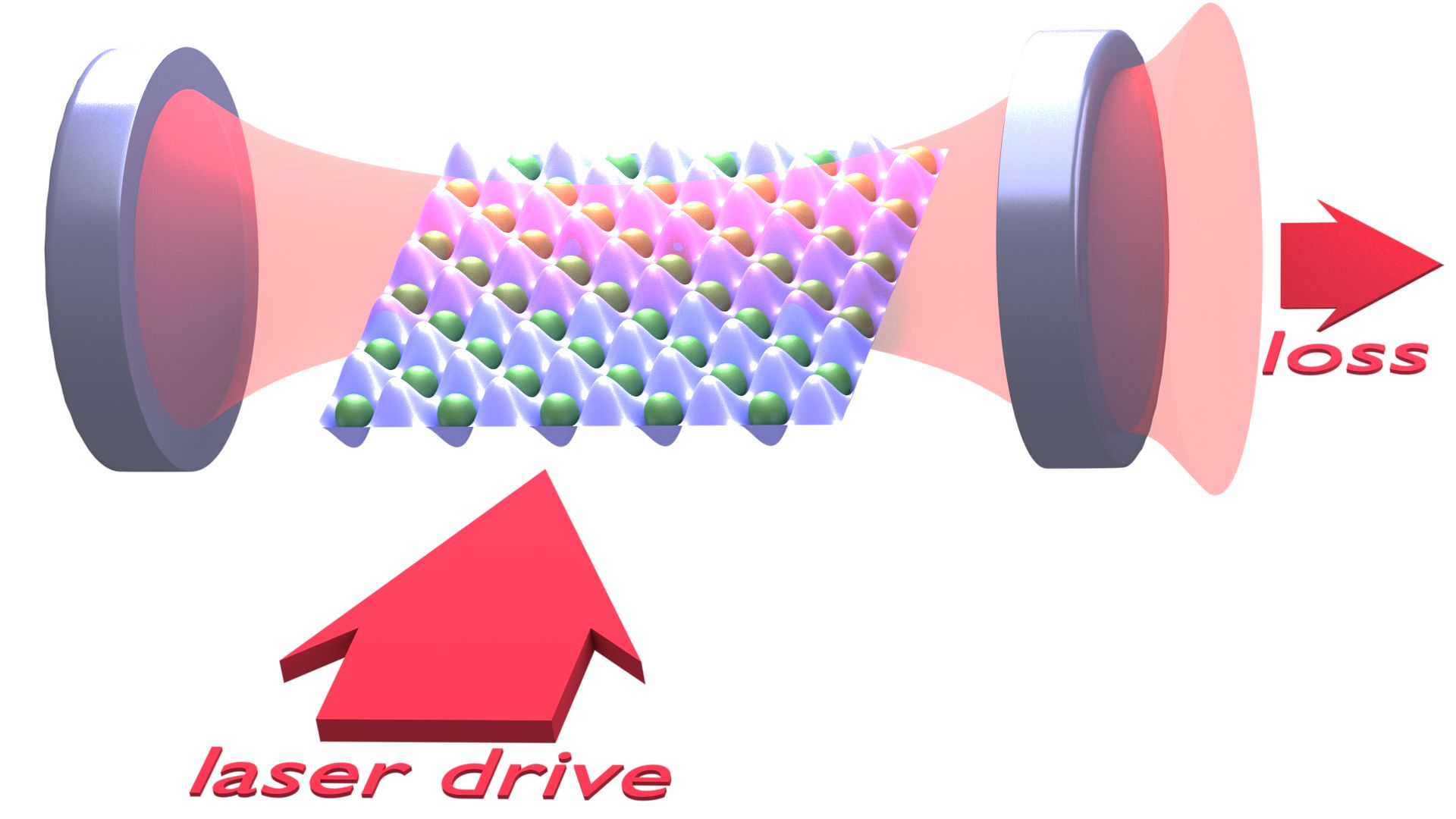}
\caption{Illustration of the coupled cavity Bose-Hubbard model setup. An atomic cloud is loaded into a square optical lattice, which is inside a single-mode high-Q Fabry-Pérot resonator. The period of the cavity mode is approximately equal to that of the optical lattice. The cavity is pumped by the side through the light scattered by the atoms from the laser drive. The system is open, photons leak out from the cavity, resulting in heating and decoherence.}
\label{fig:setup} 
\end{figure}
An atomic degenerate Bose gas trapped in an optical lattice inside a
high-finesse optical cavity gives rise to a system with competing
short- and long-range interactions. The lattice dynamics of the
ultracold atoms including on-site collisional interactions and
tunneling between adjacent sites corresponds to the usual
Bose-Hubbard-model \cite{Fisher1989Boson,Jaksch1998Cold}. The optical lattice depth set by the power of its driving laser controls the strength of these short-range effects. On top of the bosonic lattice model, there is a long-range atom-atom interaction mediated by the cavity field in photon scattering processes.  In order to be specific, we consider the geometry of the experimental setup in  Ref.~\cite{Landig2016Quantum}, illustrated in Fig~\ref{fig:setup}. A sample of bosonic atoms is placed inside a two-dimensional optical lattice, which overlaps with a mode of an optical cavity. A single mode is selected of which the wavelength is very close to that of the optical lattice. The atoms are illuminated from the side by means of a far-detuned laser source which is close to resonance with the cavity mode. The atoms scatter photons coherently between the laser and the commonly coupled cavity mode \cite{Ritsch2013Cold}. The collective coupling of the atoms to the cavity mode leads to an indirect and infinite-range interaction between the atoms, which can be controlled independently from other parameters of the lattice dynamics by means of a small detuning of the drive frequency from the cavity resonance. The effective Hamiltonian of the system reads, see Ref.~\cite{Landig2016Quantum},
\begin{subequations}
  \label{eqs:Hamiltonian}
 \begin{equation}
      H=H_\text{at}+H_\text{ph}+H_{I}, 
 \end{equation}
where
  \begin{align}
    H_\text{at}&=-J\sum_{\langle i,j\rangle} \Big (b^\dagger_i b^{}_j + b^\dagger_j b^{}_i\Big)+\frac{U_s}{2}\sum_{j} n_j (n_j-1)\,,\\
    H_\text{ph}&=-\hslash\Delta_C a^\dagger a\,,\\
    H_{I}&=\hslash \eta (a^\dagger + a)\sum_{j}(-1)^j n_j\,. \label{eq:atcav}
  \end{align}
\end{subequations}
The first term is the standard Bose--Hubbard model describing the atomic dynamics inside the optical lattice. The atomic annihilation and creation operators at site $i$ are $b_i$ and $b_i^\dagger$, respectively. The tunneling is characterised by $J$, and the strength of the on-site interaction is denoted by $U_s$. The second term represents the oscillator of the cavity field, with $a$ being the photon annihilation operator, and $\Delta_C=\omega_L-\omega_C$ the cavity detuning, i.e., the frequency difference of the pumping laser and the empty cavity.
The interaction term $H_I$ is a driving of the cavity mode which depends on the atom density $n_j=b^\dagger_j b^{}_j$, and as a key element, on a phase factor $(-1)^j$. The phase  comes from that the atoms scatter the laser light into the cavity with an amplitude depending on the  sign of the cavity mode function at the given site. This sign alternates between adjacent sites of the optical lattice, discriminating ``even'' and ``odd'' sites in a checker-board pattern. The strength of the interaction is modelled by the parameter $\eta$, which is proportional to the pumping amplitude of the laser and inversely proportional to cavity detuning $\Delta_C$.

Because of the very large detuning of the driving laser with respect to all atomic resonances, the atoms scatter photons only in a coherent manner and the spontaneous emission can be neglected. However, one must take into account that the photons leak out from the cavity through the mirrors and couple to the free-space modes. The loss process can be treated within the Markov approximation with an intensity loss rate of $2 \kappa$.  The Heisenberg equations of motion of the operators are complemented by a damping term and a corresponding Langevin-type quantum noise term. The Heisenberg--Langevin equations are 
\begin{widetext}
  \begin{subequations}
  \begin{align}
    \partial_t a &= i\hslash^{-1}[H,a]-\kappa a + \xi=(i\Delta_C-\kappa) a -i\eta \Delta n + \xi\,,\label{eq:eqmoa}\\
    \partial_t b_j &=i\hslash^{-1}[H,b_j]=i\hslash^{-1} J\sum_{\varepsilon}b_{j+\varepsilon}
     -i\hslash^{-1}U_s b^\dagger_j b^{}_j b^{}_j - i \eta (a^\dagger + a)(-1)^jb^{}_j\,,
\label{eq:eqmob}
  \end{align}
\end{subequations}
\end{widetext}
where $\partial_t$ denotes the time derivative, $\Delta n=\sum_j(-1)^j n_j$ is the imbalance between the particle numbers on the even and odd sites. It can be seen from the form of this term that the atoms act collectively on the cavity mode amplitude. In Eq.~\eqref{eq:eqmob}, the sum for $\varepsilon$ is over the 4 neighbouring sites of the site $j$. The noise term in Eq.~\eqref{eq:eqmoa} is a delta-correlated white noise with zero mean value: $\langle\xi(t)\rangle=0$.  Furthermore, let us assume that the temperature is very low compared to optical frequencies, that is, normal-ordered correlations are zero. In particular, $\langle \xi^\dagger(t') \xi(t)\rangle = 0$. All other correlations are determined by the bosonic commutation relations,
\begin{equation}
  \label{eq:xinoise}
  \big[\xi(t),\xi^\dagger(t')\big]=2\kappa \delta(t-t')\,,
\end{equation}
in second order. The goal of this paper is to reveal the dynamical consequences of this noise term during the initial short evolution time.

\section{Adiabatic dynamics of the atoms}
\label{sec:effmodel}

Since the time scale of the cavity field relaxation is the far shortest, i.e. $\kappa,\Delta_C\gg J,U_s$, the cavity mode can be slaved to the slow atomic lattice dynamics. Performing adiabatic elimination of the cavity dynamics, it results in a renormalisation of the parameters describing the atomic motion in the dynamical lattice. Furthermore, damping and decoherence are introduced into the atomic time evolution. These latter effects are modeled also by a Langevin-like noise, now as part of the atomic motion, which, in general, is not a white noise. 

Let us start by integrating out the fast cavity field equations of motion \eqref{eq:eqmoa}, with keeping the slow atomic operators constant,
\begin{subequations}
  \label{eqs:cavint}
\begin{equation}
  \label{eq:cavitysol}
  a(t)=\frac{\eta\,\Delta n}{\Delta_c+i\kappa}+\Sigma(t)\,,
\end{equation}
where the first term is the adiabatic steady state of the photon field, and the noise is
 \begin{equation}
  \label{eq:OUnoise}
  \Sigma(t)=i\int\frac{d\omega}{2\pi}\frac{\xi(\omega)e^{-i\omega t}}{\omega+\Delta_C+i\kappa}\,,
\end{equation} 
\end{subequations}
which is a  white noise filtered through the cavity mode. The commutation relation can be derived from that of the original Eq.~\eqref{eq:xinoise}, 
\begin{equation}
  \label{eq:OUcomm}
  \Big[\Sigma(t),\Sigma^\dagger(t')\Big]
  =e^{i\Delta_C(t-t')}e^{-\kappa|t-t'|}.
\end{equation}

Combining Eqs.~\eqref{eqs:cavint} with Eq.~\eqref{eq:eqmob}, we arrive at the adiabatic dynamics
  \begin{align}
    \partial_t b_j &=i\hslash^{-1} J\sum_{\varepsilon}b_{j+\varepsilon}
                     -i\hslash^{-1}U_s b^\dagger_j b^{}_j b^{}_j&\nonumber\\
                     &\qquad - i \eta \frac{2\Delta_C\,\eta\,\Delta n}{\Delta_C^2+\kappa^2}(-1)^j\, b^{}_j-i \eta\,R(t)\,(-1)^j\,b_j(t)\,,&\label{eq:eeqmob}
  \end{align}
where the last noise term includes the product of atomic operators and the noise $R(t)=\Sigma(t)+\Sigma^\dagger(t)$, being a self-adjoint operator. The mean value of $R(t)$ is zero, and the second-order correlations are evaluated similarly to Eq.~\eqref{eq:OUcomm} with the following result,
\begin{equation}
  \label{eq:Rcorr}
  \langle R(t) R(t')\rangle = \langle \Sigma(t) \Sigma^\dagger(t')\rangle
  =e^{i\Delta_C(t-t')}e^{-\kappa|t-t'|}.
\end{equation}  
When the photon decay $\kappa^{-1}$ is much shorter than the other timescales of the problem, we can approximate Eqs.~\eqref{eq:OUcomm} and \eqref{eq:Rcorr} by a delta-correlated noise,
\begin{equation}
  \label{eq:OUcommapp}
  \langle R(t) R(t')\rangle \approx D_R \,\delta(t-t')\,.
\end{equation}
with
\begin{equation}
D_R = \frac{2\kappa}{\Delta_C^2+\kappa^2}.
\label{eq:DRdef}
\end{equation}
This approximation corresponds to the broad bandwidth reservoir assumption used for Markovian decay. 

We note that the appearance of fluctuations due to the lossy cavity mode is accompanied by non-adiabatic drift terms, the so-called cavity cooling or heating, depending on the sign of the detuning $\Delta_C$ Ref.~\cite{Nagy2009Nonlinear}. However, this is beyond the adiabatic approximation and can be safely neglected in the limit $\kappa \gg \omega_R$.

\section{Manifestation of the multiplicative noise in the different phases}
\label{sec:noiseeff}

In this section we analyze the effects of the noise created by the leakage of cavity photons on the atomic dynamics. We assume that this effect is small and that the noise can be considered as an additional effect on top of the quantum fluctuations arising from the interactions within the many-body system.

The study has to be done separately for the distinct cases corresponding to the different possible thermodynamic phases \cite{Chen2016Quantum,Dogra2016Phase,Niederle2016Ultracold,Sundar2016Lattice,Liao2018Theoretical,Panas2017Spectral,Flottat2017Phase}. In fact, the model allows for four different ground state phases depending on the preservation or breaking of two different symmetries. Firstly, the global U(1) symmetry associated with particle number conservation, the breaking of which corresponds to the presence of off-diagonal long-range order. Secondly, the $\Z_2$ symmetry of even and odd lattice sites, which the atom--cavity interaction Eq.~\eqref{eq:atcav} can break spontaneously. When the $\Z_2$ symmetry is not broken, all averages on the even and odd lattice sites are equal. When the $\Z_2$ symmetry is broken, the expectation values of operators on the even and on the odd sublattice can be different. The four phases are: the \emph{superfluid phase} with broken U(1) and unbroken $\Z_2$ symmetry, the \emph{supersolid phase} with both symmetries broken, the \emph{Mott phase} with both phases unbroken, and finally, the \emph{charge-density wave (CDW) phase} with a broken $\Z_2$ and unbroken U(1) symmetry.

The broken U(1) symmetry phases emerge when the on-site atom-atom interaction is small compared to tunneling, $U_s\ll J$. In this weakly-interacting limit the equations of motion of the atomic field operators can be truncated at second order in the usual perturbative way, and can be eventually dealt with Bogoliubov transformation \cite{Castin2001BoseEinstein}. In the other limit, when $U_s\gg J$, one goes to the strongly interacting phases, where U(1) is unbroken. Here the Mott phase and the CDW phase are described after a canonical transformation to the low energy degrees of freedom \cite{Altman2002Oscillating,Huber2007Dynamical}.

\subsection{Superfluid and supersolid phases}
\label{ssec:noisesf}

\begin{figure}[tb!]
\centering
\includegraphics{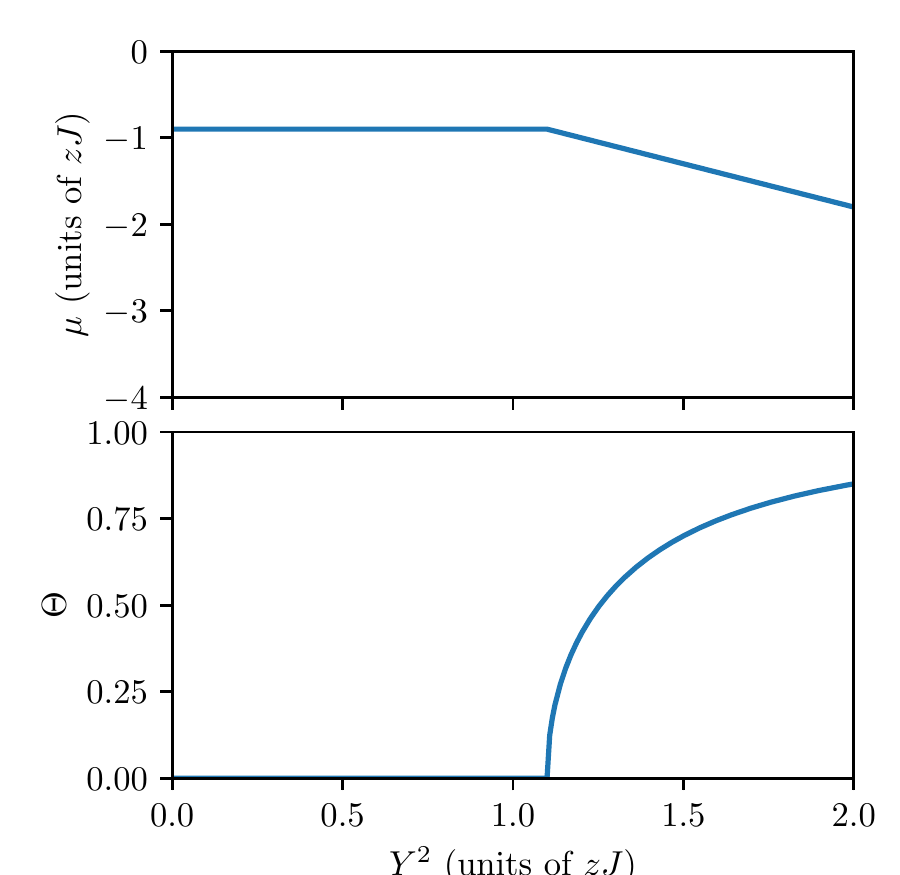}
\caption{The chemical potential and the order parameters as a function of the pump power $Y^2$. We measure the chemical potential and $Y^2$ in units of $z J$, and use $g=0.1 zJ$.}
\label{fig:sfmf}
\end{figure}

In the weakly interacting limit, where $U_s\ll J$, Bose-Einstein condensation occurs for sufficiently low temperatures. The wavefunction of the condensate is proportional to the expectation value of the atomic ladder operators, where the coefficient is the square root of the average atomic density of the condensate. Close to zero temperature almost all the atoms are condensed, and the condensate density can be approximated by the total density. It is convenient to separate the mean values from the fluctuations,
\begin{equation}
\label{eq:bdecoup}
b_i(t)=\left(\sqrt{n}\beta_i+\delta b_i(t)\right)e^{-i\mu t/\hslash},
\end{equation}
where $n=N/M$ is the total density of atoms, i.e. the total number of atoms $N$ divided by the total number of sites $M$. Furthermore, $\mu$ plays the role of the chemical potential. When inserting Eq.~\eqref{eq:bdecoup} into Eq.~\eqref{eq:eeqmob} and collecting the terms proportional to $\sqrt{n}$, we arrive at the Gross-Pitaevskii equation (GPE),
\begin{equation}
0 = -J\sum_{\varepsilon}\beta_{j+\varepsilon} -\mu \beta_j  - (-1)^jY^2\Theta\beta_j+g|\beta_j|^2\beta_j.
\label{eq:GPE}
\end{equation}
We introduced the dimensionless parameter $Y^2=-2\hslash\Delta_C\eta^2 N/(\Delta_C^2+\kappa^2)$ characterising the power of the pumping laser and the scaled on-site interaction strength $g=nU_s$. Furthermore, we introduced the supersolid order parameter $\Theta$, which is the difference between the even and odd site condensate densities, which are normalised to unity in agreement with Eq.~\eqref{eq:bdecoup},
\begin{align}
\Theta&=\frac{1}{M}\sum_j (-1)^j |\beta_j|^2,\label{eq:thetadef}\\
1&=\frac{1}{M}\sum_j |\beta_j|^2.\label{eq:condnorm}
\end{align}
Equations~\eqref{eq:GPE},~\eqref{eq:thetadef},~\eqref{eq:condnorm} form a closed set for the chemical potential $\mu$, the supersolid order parameter $\Theta$ and the condensate wavefunction $\beta_j$. They have a remarkably simple solution both in the superfluid and in the supersolid phases. In both phases,
\begin{equation}
\Theta=\sqrt{1-\frac{z^2 J^2}{(g-\mu)^2}},
\label{eq:Thetamu}
\end{equation}
where $z$ is the coordination number of the lattice. For the square lattice, $z=4$. Furthermore, by introducing $\beta_e=\sqrt{1+\Theta}$ for the even sites and $\beta_o=\sqrt{1-\Theta}$ for the odd sites, the solution becomes
\begin{equation}
\beta_j=
\begin{cases}
\beta_e,&\text{for }j\text{ even},\\
\beta_o,&\text{for }j\text{ odd}.
\end{cases}
\end{equation}
Finally, the following equation has to be fulfilled:
\begin{equation}
0=\Theta\,(\mu+Y^2-2g),
\label{eq:finmfeq}
\end{equation}
from which either $\Theta=0$, or $\mu=2g-Y^2$. The superfluid phase is characterised by $\Theta=0$, i.e., with a homogeneous condensate $\beta_e=\beta_o=1$. In this phase, from Eq.~\eqref{eq:Thetamu}, we get $\mu=g-zJ$. In the supersolid phase $\mu=2g-Y^2$, and from Eq.~\eqref{eq:Thetamu}, we arrive to $\Theta=[1-z^2 J^2/(Y^2-g)^2]^{1/2}$. The critical pumping power separating the two phases is at
\begin{equation}
\label{eq:threshold}
Y_c^2=z J + g.
\end{equation}
The chemical potential and the order parameters are plotted in Fig.~\ref{fig:sfmf}.

Using again the substitution Eq.~\eqref{eq:bdecoup} in Eq.~\eqref{eq:eeqmob} but now keeping the terms linear in the atomic operators and the noise, we get the equation of motion for the fluctuations,
\begin{widetext}
\begin{equation}
i\hslash\partial_t\delta b_j=-J\sum_{\varepsilon} \delta b_{j+\varepsilon} -\mu \delta b_j - (-1)^j Y^2 \left[\Theta\delta b_j + \frac{\beta_j}{M}\,\sum_k(-1)^k\beta_k(\delta b_k + \delta b_k^\dagger)\right] + g \beta_j^2 (2\delta b_j + \delta b_j^\dagger) + (-1)^j\hslash \sqrt{n} \eta  R(t) \beta_j.
\label{eq:flucteqsf}
\end{equation}
We restrict ourselves to the low energy excitations and neglect the wavenumber dependence of the fluctuations. That is, we keep only the two relevant modes of the symmetry breaking: the fluctuations of the ladder operators of the even and odd sites. The 2 coupled equations become
\begin{subequations}
\label{eqs:flucttwo}
\begin{align}
i\hslash\partial_t\delta b_e &= [(g-\mu)+\Theta(g-Y^2)]\delta b_e -J z \delta b_o + \beta_e^2 (g - \frac{1}{2}Y^2)(\delta b_e + \delta b_e^\dagger)
+\frac{1}{2}Y^2\beta_e\beta_o(\delta b_o + \delta b_o^\dagger)+\sqrt{n}\hslash\eta\beta_e R(t),\\
i\hslash\partial_t\delta b_o &= -J z \delta b_e + [(g-\mu)-\Theta(g-Y^2)]\delta b_o + \frac{1}{2}Y^2\beta_e\beta_o(\delta b_e + \delta b_e^\dagger)+ \beta_o^2 (g - \frac{1}{2}Y^2)(\delta b_o + \delta b_o^\dagger)
-\sqrt{n}\hslash\eta\beta_o R(t).
\end{align}
\end{subequations}
\end{widetext}
These Bogoliubov equations are constant-coefficient inhomogeneous linear
differential equations with a Langevin-type noise added as driving. Another 2 equations describing the time evolution of $\delta b_e^\dagger$ and $\delta b_o^\dagger$ have to be added to get a closed set of equations for the fluctuations. These latter two equations are obtained by taking the hermitian conjugates of Eqs.~\eqref{eqs:flucttwo}. First, let us gather the fluctuation operators to the formal vector $\mathbf{w}=(\delta b_e,\delta b_o,\delta b_e^\dagger, \delta b_o^\dagger)$. Then, the four Bogoliubov equations become
\begin{equation}
i\hslash\partial_t \mathbf{w} = M_{4\times 4}\,\mathbf{w} + \mathbf{\Xi}(t),
\label{eq:bogoeq}
\end{equation}
where the $4\times4$ coefficient matrix $M_{4\times 4}$ and the noise vector $\mathbf{\Xi}$ can be read off from Eqs.~\eqref{eqs:flucttwo}. The solution is obtained by finding the normal modes, i.e., diagonalising the coefficient matrix. The eigenvalues of $M_{4\times 4}$ come in $\pm$ pairs, and 2 of the 4 eigenvalues are identically zero in the entire Bose condensed phase \cite{Castin2001BoseEinstein}. The normal mode with zero eigenvalue, which we call the zero mode, corresponds to phase fluctuations of the condensate. The corresponding eigenvector is proportional to the condensate wavefunction, and therefore, the zero mode does not describe any of the fluctuations orthogonal to the condensate. Furthermore, its operator is anti-hermitian and is decoupled from the dynamics of the other normal mode with eigenvalues $\pm\lambda$. This latter excitation describes density waves with a period of 2 lattice sites. At the transition point between the superfluid and the supersolid phases, this excitation also becomes soft, therefore, we call it as the soft mode. In other words, we introduce $\delta b_z$ and $\delta b_s$, corresponding to the zero and soft modes, respectively, as
\begin{equation}
(\delta b_e,\delta b_o)=\bm{\gamma}\, \delta b_z + \bm{\varepsilon}\, \delta b_s,
\label{eq:zeromode}
\end{equation}
where $\bm{\gamma}=(\beta_e,\beta_o)/\sqrt{2}$, and
$\bm{\varepsilon}=(\beta_o,-\beta_e)/\sqrt{2}$, orthogonal vectors. These two vectors span the 2-dimensional space, from which, only the direction orthogonal to the condensate, i.e., the soft mode is relevant for us. The soft mode decouples from the purely anti-hermitian zero mode, and its
Bogoliubov equation is obtained directly from Eq.~\eqref{eq:bogoeq}
simply by taking the appropriate submatrix of $M_{4\times 4}$ and subvector of $\mathbf{\Xi}$,
\begin{equation}
  i\hslash\partial_t \mathbf{v} = M\,\mathbf{v} + \mathbf{\Xi}^{'}(t)\,,
\label{eq:softm_fluct}
\end{equation}
with the reduced vector $\mathbf{v}=(\delta b_s,\delta b_s^\dagger)^T$
and noise $\mathbf{\Xi}^{'}(t) = (\zeta, -\zeta)^T$.
The $2\times 2$ coefficient matrix reads
\begin{widetext}
\begin{equation}
M = \left(\begin{matrix}
2(g-\mu)+\frac{z^2 J^2}{(g-\mu)^2}(g-Y^2) & \frac{z^2 J^2}{(g-\mu)^2}(g-Y^2)\\
\frac{-z^2 J^2}{(g-\mu)^2}(g-Y^2) & -2(g-\mu)-\frac{z^2 J^2}{(g-\mu)^2}(g-Y^2)
\end{matrix}\right)\,,
\end{equation}
\end{widetext}
while the effective noise operator is
\begin{equation}
\zeta = \sqrt{2 n}\hslash\eta\frac{z\,J}{g-\mu}R(t)\,.
\end{equation}
Note, that this noise operator inherits the properties of $R(t)$, hence it is Hermitian and its correlation is determined by Eq.~\eqref{eq:OUcommapp}.

The noise leads to a diffusion-like process, that depletes the superfluid or supersolid ground states by heating the condensate atoms into the orthogonal fluctuation mode. In the following we calculate the rate by which the atoms leave the Bose condensed state. The matrix $M$ is diagonalised by its real right and left
eigenvectors, $M \mathbf{r}^{(k)} = \omega_k \mathbf{r}^{(k)}$ and $(\mathbf{l}^{(k)})^T
M = \omega_k (\mathbf{l}^{(k)})^T$. Their scalar product is conveniently
normalised, $(\mathbf{l}^{(k)}, \mathbf{r}^{(k)}) = 1$. Multiplying
Eq.~(\ref{eq:softm_fluct}) from the left with $(\mathbf{l}^{(k)})^T$, we obtain
the equation of motion of the normal modes $\rho_k = (\mathbf{l}^{(k)}, \mathbf{v})$
that reads
\begin{equation}
  i\hslash\partial_t \rho_k = \hslash\omega_k\rho_k + Q_k \,,
\label{eq:normal_modes}
\end{equation}
where $Q_k(t) = (\mathbf{l}^{(k)}, \mathbf{\Xi}'(t))$ is the projection of the noise
vector to each mode.

\begin{figure}[bt!]
\centering
\includegraphics{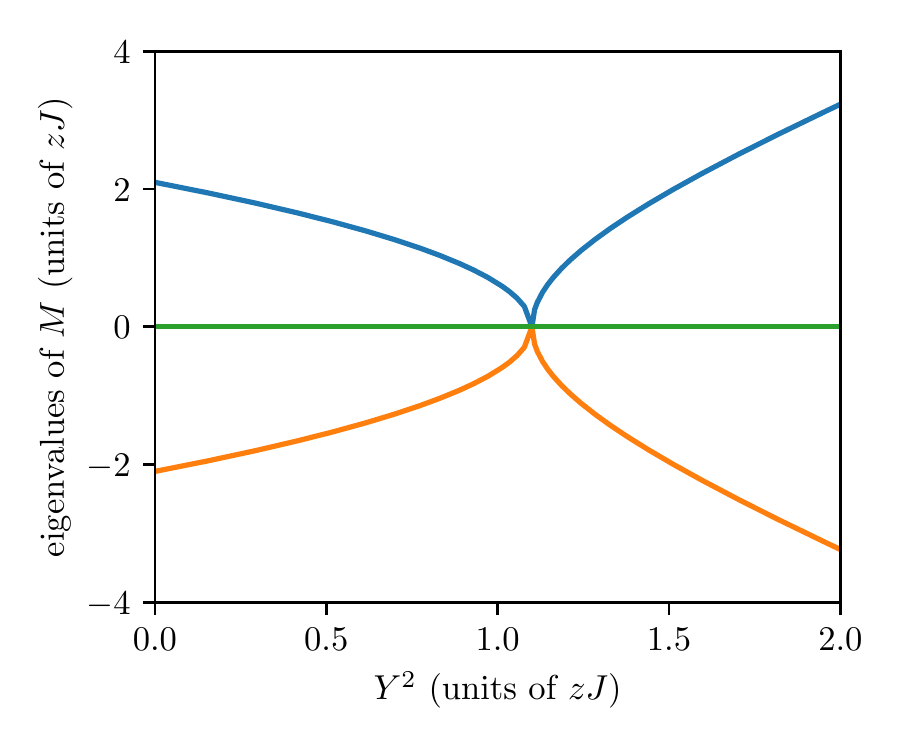}
\caption{The eigenvalues of the Bogoliubov problem as a function of the pumping power $Y^2$.}
\label{fig:evals}
\end{figure}

The normal mode frequencies (eigenvalues of the coefficient matrix $M$) are $\omega_{1,2} = \pm \hslash^{-1}\lambda$, with
\begin{equation}
\lambda=2\sqrt{(g-\mu)^2+z^2J^2\frac{g-Y^2}{g-\mu}}.
\label{eq:softmodeeval}
\end{equation}
The eigenvalues are plotted in Fig.~\ref{fig:evals}. The corresponding
normal modes form a Hermitian adjoint pair $\rho_2 =
\rho_1^{\dagger}$. Their correlation function can be directly calculated
from the formal solution of Eq.~(\ref{eq:normal_modes}),
\begin{multline}
  \langle \rho_k(t) \rho_l(t) \rangle =
  (\rho_k(0)\rho_l(0))e^{-i(\omega_k + \omega_l)t} \\
  - D_R 2n \eta^2 \frac{z^2J^2}{(g-\mu)^2} \frac{1 - e^{-i(\omega_k +
      \omega_l)t}}{i(\omega_k + \omega_l)} \sum_{i,j} (-1)^{i+j} l_i^{(k)}
  l_j^{(l)}.
\label{eq:normal_mode_correlation}
\end{multline}

The fluctuation number is expressed with the normal modes
\begin{equation}
  \langle \delta b_s^\dagger \delta b_s \rangle = \sum_{k,l}
  \langle \rho_k \rho_l \rangle r_2^{(k)} r_1^{(l)} \,.
\label{eq:fluct_num}
\end{equation}

\begin{figure}[bt!]
\centering
\includegraphics{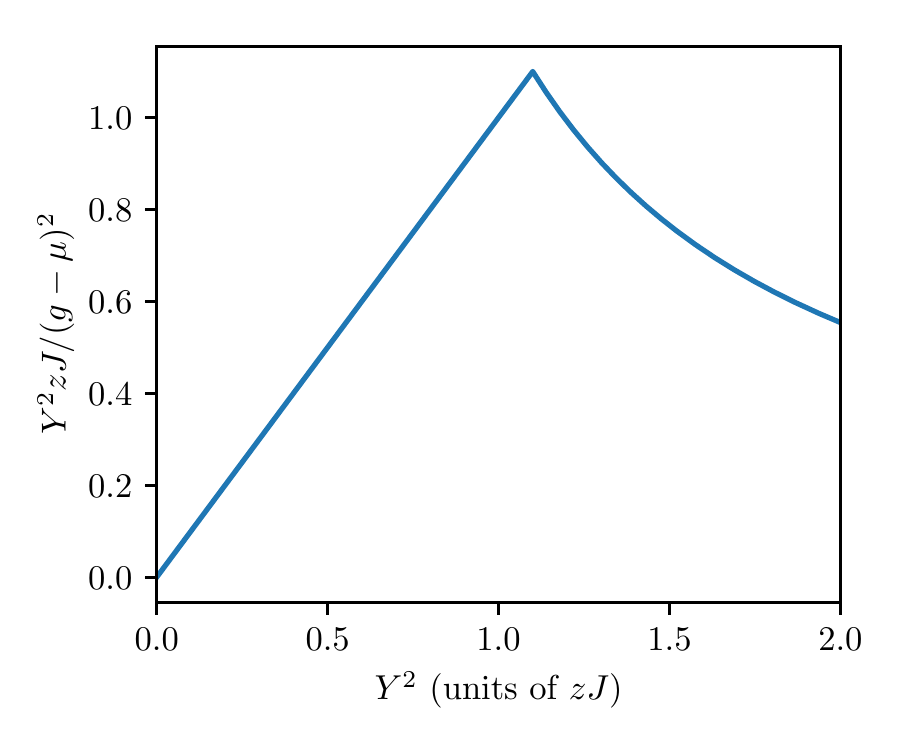}
\caption{The effective diffusion rate Eq.~\eqref{eq:difffin}, by which the atoms leave the superfluid and supersolid ground states. More precisely, we plot the dimensionless quantity $Y^2 zJ/(g-\mu)^2$ as a function $Y^2$. The diffusion rate exhibits a cusp at the transition point between the superfluid and supersolid phases.}
\label{fig:diffusion} 
\end{figure}

The summation over the eigenvector components gives a
factor of $-1$. According to Eq.~\eqref{eq:normal_mode_correlation}, the system obeys an exponential relaxation law towards the steady state when the eigenvalue $\lambda$ has an imaginary part, such as in Ref.~\cite{Szirmai2010Quantum}. This finding is similar to what was found in Ref.~\cite{Chiacchio2018Tuning} in the bad-cavity limit and for the case of vanishing on-site interactions. For short times, $t \ll \lambda^{-1}$ the incoherent
population, Eq.~(\ref{eq:fluct_num}) builds up linearly in time,
hence the rate characterising the diffusion is
\begin{equation}
\label{eq:difffin}
  \frac{\langle \delta b_s^\dagger \delta b_s \rangle}{t} = D_R 2n
  \eta^2 \frac{z^2J^2}{(g-\mu)^2} = \frac{2\kappa z J}{M \hslash|\Delta_C|}\,\frac{Y^2 zJ}{(g-\mu)^2}\,.
\end{equation}
The right-hand side of Eq.~\eqref{eq:difffin} is written as a product of two factors. The first one sets the dimension and the order of magnitude of the diffusion, while the second factor is dimensionless and is on the order of unity. This latter quantity is plotted in Fig~\ref{fig:diffusion}. In the superfluid phase it is proportional to $Y^2$, while in the supersolid phase it goes to zero according to $1/Y^2$. The first factor has the dimension of $s^{-1}$, and is combined from 3 different frequency scales: the photon escape rate ($\kappa$), the magnitude of the cavity detuning ($|\Delta_C|$) and the optical-lattice tunneling rate ($z J/\hslash$). In the experiments, the first two of the three frequencies are usually chosen to be close to each other, while $z J$ is much smaller then the other two. Furthermore, the number of lattice sites illuminated by the cavity mode is on the order of a thousand ($M\sim10^3$). Therefore, we estimate $2\kappa zJ/(M\hslash |\Delta_C|)\sim 10^{-2}zJ \text{ --- } 10^{-3}zJ$. Clearly, this magnitude of the diffusion process allows for considering the tunneling dynamics to be coherent on a long enough time scale. Notably, some of the technical noise sources can be more restrictive in this respect. However, one must be aware of that, with the cavity detuning $\Delta_C$ going to zero, the photon outcoupling process can generate a large number of photons within the cavity.  These results are in complete analogy to the model without the additional optical lattice \cite{Nagy2010DickeModel}; the noise leads to the departure of the system from its quantum ground state, and for long times it relaxes exponentially to a steady state with an excess noise depletion of the condensate \cite{Szirmai2009Excess} in the cavity-cooling regime, while it relaxes exponentially to the infinite-temperature state in the cavity-heating case. 

\subsection{Mott and CDW phases}
\label{ssec:noiseMott}

When the strength of the on-site repulsion is much larger than the amplitude of the hopping ($U\gg J$), multiple occupations of a site become energetically penalised. For commensurate fillings, i.e., when the number of atoms is integer times the number of lattice sites, each lattice site becomes populated with  exactly the same number of atoms. In the same time, particle number fluctuations become suppressed at each individual site. Depending on the strength of the transverse laser driving, the system can be in a Mott state or in a CDW state. For weak pumping, the Mott state is realised, where each site has the same number of particles. In contrast, when the pumping strength is large, the CDW state is the ground state, where all even sites have the same occupation and all odd sites have the same occupation, but these two are different. Both in the Mott- and CDW phases, the low energy excitations are no longer the Bogoliubov quasiparticles but rather particles and holes of a strongly correlated system. These quasiparticle excitations have to be introduced separately for each phase. To be specific, we are going to study the excitations over the Mott state with one particle per site and over the CDW state with one particle per two sites. The generalisation of the theory to other Mott- and CDW states is straightforward and the conclusions do not change qualitatively. For a more transparent presentation, we also completely neglect the effect of the kinetic energy, which can be reintroduced with the help of perturbation theory, as was done for the Mott-superfluid transition in Refs.~\cite{Altman2002Oscillating,Huber2007Dynamical}.

Without tunneling, the Hamiltonian~\eqref{eqs:Hamiltonian} becomes
\begin{multline}
\label{eq:Hamprob}
H=H_\text{ph}+\sum_{j}\bigg[\frac{U_s}{2}n_j(n_j-1)-\mu n_j\\
+\hslash \eta (a^\dagger + a)(-1)^j n_j\bigg].
\end{multline}
In the one particle per site Mott lobe, we follow the route of Ref.~\cite{Altman2002Oscillating} and truncate the single site Hilbert space for the three lowest occupied states, $|0\rangle_j$, $|1\rangle_j$, $|2\rangle_j$. We introduce three Schwinger boson operators creating these states,
\begin{subequations}
\label{eqs:SchwingerBosons}
\begin{align}
|1\rangle_j&=t_{1,j}^\dagger\vac,\label{eq:t1}\\
|0\rangle_j&=t_{0,j}^\dagger\vac,\label{eq:t0}\\
|2\rangle_j&=t_{2,j}^\dagger\vac.\label{eq:t2}
\end{align}
\end{subequations}
The state $\vac$ is a fictitious vacuum state, not present in our original Hilbert space. In order to exclude the unphysical states, these Schwinger bosons must share a single excitation at each site, therefore we have the constraint
\begin{equation}
t_{0,j}^\dagger t^{}_{0,j} + t_{1,j}^\dagger t^{}_{1,j} + t_{2,j}^\dagger t^{}_{2,j} = 1.
\label{eq:constraint}
\end{equation}
In the present case, the ground state is a tensor product state of the singly occupied sites,
\begin{equation}
|GS\rangle=\prod_j|1\rangle_j.
\label{eq:MottGS}
\end{equation}
That is, in the ground state, the occupation of the $t_1$ boson is 1 at each site, while the other bosons have zero occupations. Low energy excited states have these other bosonic excitations mixed in with some amplitude much smaller than unity, while the $t_{1,j}$ bosons having still an occupation very close to unity. Thus, by using Eq.~\eqref{eq:constraint}, the Hamiltonian can be written as
\begin{equation}
H=H_\text{ph}+\sum_j\left[\tilde \mu_j t_{0,j}^\dagger t^{}_{0,j}+(U-\tilde \mu_j)t_{2,j}^\dagger t^{}_{2,j}-\tilde \mu_j\right],
\label{eq:Ham_w_t0_t2}
\end{equation} 
with
\begin{equation}
\tilde \mu_j=\mu-\hslash\eta(-1)^j(a^\dagger+a).
\label{eq:localmu}
\end{equation}
Equation \eqref{eq:Ham_w_t0_t2} is quadratic in the excitations, i.e., both the hole-type excitations created by $t^\dagger_{0,j}$ and the particle-like excitations of $t^\dagger_{2,j}$. In the Mott phase, the effect of the photon field is through renormalising the (chemical) potential. Integrating out the photon field, we still have
Eqs.~\eqref{eqs:cavint}, but now with
\begin{equation}
\Delta n = \sum_j (-1)^j \left( 1 - t^\dagger_{0,j} t^{}_{0,j} +  t^\dagger_{2,j} t^{}_{2,j}\right).
\label{eq:partno_Mott}
\end{equation}
To lowest order in the single particle per site Mott phase, we can approximate $\Delta n\approx\sum_j (-1)^j=0$. That is, in the Mott phase there is no classical part of the cavity field $a_{\text{ss}}=0$. Therefore, the chemical potential is renormalised only by a noise,
\begin{equation}
\tilde \mu_j=\mu-\hslash\eta(-1)^j R(t).
\label{eq:chempotnoise}
\end{equation}
Therefore, in the Mott phase, for zero tunneling, the effect of cavity decay is to make
the chemical potential noisy. As the Mott phase is gapped, there is no
effect of the noise on the ground state while we stay away from the
phase boundary. The fluctuation of the chemical potential  affects, on the other hand,
the particle- and hole-like excitations of the Mott phase: 
the corresponding quasiparticle resonances get broadened by an amount of
\begin{equation}
\Delta = \frac{2 \, \kappa \,\eta^2}{\kappa^2 + \Delta_C^2} \,. 
\end{equation}
When a small enough tunneling is introduced, the situation can be more complicated, as the noise can dress also the quasiparticles. This effect must be small, as it has to vanish for zero tunneling. In Ref.~\cite{Chiacchio2018Tuning} the authors found anomalous and normal diffusion towards the steady state similar to the standard optical lattice Mott insulator \cite{Poletti2013Emergence}.

The present analysis was performed for the first Mott sate, i.e., the one with $n=1$ particles per site. This analysis can simply be generalised to any of the Mott states with $n$ particles per site by keeping the most relevant three states, namely $|n-1\rangle_j$, $|n\rangle_j$, and $|n+1\rangle_j$ at each site. Equations through \eqref{eq:MottGS}---\eqref{eq:partno_Mott} has to be modified, although in a straightforward manner. Nevertheless, the final result for the noisy chemical potential, Eq.~\eqref{eq:chempotnoise}, holds for each Mott state.

In order to generalise the previous analysis further to the CDW phase, we need to introduce two sublattices in a checkerboard setting. One sublattice has $n_e$ particles on each site, while on the other sublattice each site has $n_o$ atoms. We shall refer to these sublattices as the even and odd sublattice, respectively. A unit cell now contains two neighbouring sites, each from a different sublattice. The Hamiltonian  reads
\begin{multline}
\label{eq:HamCDWunit}
H=H_\text{ph}+\sum_m\bigg\lbrace \frac{U_s}{2}\left[n_{m,e} (n_{m,e} - 1) + n_{m,o}(n_{m,o} - 1)\right]\\
-\mu(n_{m,e}+n_{m,o})+\hslash\eta(a + a^\dagger)(n_{m,e}-n_{m,o})\bigg\rbrace.
\end{multline}
The sum goes over the unit cells indexed by $m$. The number of unit cells is half the number of sites. For simplicity we work with the $n_e=1$ and $n_o=0$ CDW phase. As the occupation number of each site on the even sublattice is unity, we use the same 3 states and the same 3 Schwinger bosons as in Eqs.~\eqref{eqs:SchwingerBosons}. On the odd sublattice, we have no atoms at all in the ground state, and we use only two Schwinger bosons corresponding to the empty and to the singly occupied sites.
\begin{subequations}
\label{eqs:SBs}
That is, if site $j$ is the even site in unit cell $m$,
\begin{equation}
|1\rangle_{m,e}=t_{1,m}^\dagger\vac,\ |0\rangle_{m,e}=t_{0,m}^\dagger\vac,\ |2\rangle_{m,e}=t_{2,m}^\dagger\vac.\label{eq:SBt}
\end{equation}
While, if $j$ is on the odd sublattice inside unit cell $m$,
\begin{equation}
|0\rangle_m=s^\dagger_{0,m}\vac,\quad|1\rangle_m=s^\dagger_{1,m}\vac.\label{eq:SBs}
\end{equation}
\end{subequations}
We have two constraints for each unit cell, namely,
\begin{subequations}
\begin{align}
t_{0,m}^\dagger t^{}_{0,m} + t_{1,m}^\dagger t^{}_{1,m} + t_{2,m}^\dagger t^{}_{2,m} = 1,\label{eq:constt}\\
s_{0,m}^\dagger s^{}_{0,m} + s_{1,m}^\dagger s^{}_{1,m}  = 1.\label{eq:consts}
\end{align}
\label{eqs:CDWSchwinger}
\end{subequations}
Now Eq.~\eqref{eq:HamCDWunit} is expressed with the new Schwinger bosons, also using the constraints Eqs.~\eqref{eqs:CDWSchwinger},
\begin{multline}
\label{eq:HamCDWSchwinger}
H=H_\text{ph}+\sum_m\big\lbrace
-(\mu+\Delta\mu)s_{1,m}^\dagger s^{}_{1,m} + (\mu-\Delta\mu) t^\dagger_{0,m}t^{}_{0,m} + \\
[U_s-(\mu-\Delta\mu)] t^\dagger_{2,m}t^{}_{2,m} - (\mu-\Delta\mu) \big\rbrace,
\end{multline}
with
\begin{equation}
\Delta\mu=\hslash \eta (a + a^\dagger)=\frac{2\hslash\eta^2\Delta_C}{\Delta_C^2+\kappa^2}\Delta n + \hslash\eta R.
\label{eq:deltamu}
\end{equation}
Where the photon field is adiabatically eliminated in order to reach the final result. In this phase, $\Delta n = N$, since the odd sites are empty, and $\Delta\mu$ is negative (note, that $\Delta_C$ is negative). In fact, this CDW phase is stable while the energy of the $s_{1,m}$ particle excitations on the odd sites is positive, i.e., while $|\Delta\mu|>\mu$. Below that, the system relaxes to the $n=1$ Mott phase. Notice, that the noise term randomly pulls $\Delta\mu$, and the boundary of the phase gets smoothed out.

\section{Discussion and summary}
\label{sec:sum}

In this paper, we studied the dynamics of a lattice Bose gas, which apart from the external optical lattice, is immersed in the optical field of a single-mode high-Q Fabry-Pérot interferometer. The wavelength of the resonator is tuned close to that of the optical lattice in order to allow the lattice gas to form a grating. This atomic grating can scatter light into the cavity mode either with constructive interference, thus allowing a classical cavity field to build up, or scatter the light destructively thereby destroying the cavity field. There is a phase transition in between these two regimes separating two phases with different lattice periodicity.

We derived analytical expressions for the superfluid--supersolid phase transition in the weak-coupling limit, and an analytical formula for the excitation energy of the critical density wave. 
In the experimental realisation of the cavity Bose--Hubbard model, the
optical resonator is an inherently open quantum system. Photons from
the laser pump are scattered into the cavity, and finally leak out through the
cavity mirrors. The corresponding dissipation process leads to a
quantum noise, that can excite the system out from its ground
states. It is a fundamental question what time limitation this intrinsic quantum process imposes on the validity of the ground state description, which is substantial to all the studies relying on a Hamiltonian description of the cavity Bose--Hubbard model.  We calculated analytically the effects of photon loss dissipation in the
superfluid and Mott phases. We showed that the dissipative quantum
noise has substantially different effects in the two phases. In the
superfluid phase, the noise heats the system out from its ground state
with a time scale ${2\kappa z J}/({M \hslash|\Delta_C|})$. This time scale is a slow one except for very small detunings.

In the Mott-type phase the gap protects the population in the ground state against the photon field fluctuation noise and allows only a much slower approach towards equilibrium. However, the noise blurs also the phase boundaries and affects the excitation spectrum by the quasiparticle resonance broadening with an amount of ${2 \, \kappa \,\eta^2}/({\kappa^2 + \Delta_C^2})$.

\section*{Acknowledgements}

This work was supported by the National Research, Development and Innovation Office of Hungary (NKFIH) within the Quantum Technology National Excellence Program  (Project No. 2017-1.2.1-NKP-2017-00001) and by Grant No. K115624. D. Nagy was supported  by the J\'anos Bolyai Fellowship of the Hungarian Academy of Sciences.

\bibliography{beccav}

\end{document}